\documentclass[article,showpacs,showkeys,preprintnumbers,amsmath,amssymb,12pt]{revtex4}
\usepackage{graphicx}% Include figure files
\usepackage{dcolumn}% Align table columns on decimal point
\usepackage{bm}% bold math
\usepackage{slashed} %for stylis feynmann slash
\usepackage[utf8]{inputenc}
\usepackage{ amssymb } % for math
\newcommand{\Lagr}{\mathcal{L}}

\newcommand{\dd}{\textrm{d}}

\begin{document}

\preprint{APS/123-QED}

\title{Studying the Landau mass parameter of the extended $\sigma$-$\omega$ model for neutron star matter}% Force line breaks with \\

\author{David Alvarez-Castillo}
 \email{alvarez@theor.jinr.ru}
 \affiliation{
 H.Niewodnicza\'nski Institute of Nuclear Physics, Radzikowskiego 152, 31-342 Krak\'ow, Poland \\
 Bogoliubov Laboratory for Theoretical Physics, Joint Institute for Nuclear Research, 6 Joliot-Curie Street, 141980 Dubna, Russia} %Lines break automatically or can be forced with \\
%\affiliation{%
\author{Alexander Ayriyan}
\affiliation{
Laboratory for Information Technologies, Joint Institute for Nuclear Research, 6 Joliot-Curie street, 141980 Dubna, Russia \\
Computational Physics and IT Division, A.I. Alikhanyan National Science Laboratory, Armenia
}%

\author{Gergely G\'abor Barnaf\"{o}ldi and P\'eter P\'osfay}
 \affiliation{
Department for Theoretical Physics, Wigner Research Centre for Physics, 29-33 Konkoly-Thege Mikl\'os Street, H-1121 Budapest, Hungary
}%

\date{\today}% It is always \today, today,

\begin{abstract}
We present a Bayesian analysis of the Landau mass within the extended $\sigma$-$\omega$ model for neutron star matter. To this purpose, we consider the mass measurement of the object PSR 0740+6620, the tidal deformability estimation from the GW170817 and the mass-radius estimate of PSR J0030+0451 by NICER. Using Landau mass as free parameter of the theory, we rely on the prediction power of the Bayesian method to find the best value for this nuclear quantity. 
\end{abstract}

\pacs{Valid PACS appear here}% PACS, the Physics and Astronomy
%Use showkeys class option if keyword
\maketitle
%%%%%%%%%%%%%%%%%%%%%%%%%%%%%%%%%%%%%%%%%%%%%%%%%%%%%%%%%%%%%%%%%%%%%%%%%%%%%%%%%%%%%%%%%%%%%%%%%%
\section{Introduction}
\label{sec:intro}

The study of dense nuclear matter is an active research area. Experiments carried on relativistic collisions of heavy ions can provide estimates on the properties of hot, dense matter. On the other hand, celestial objects like neutrons stars bear the densest type of nuclear matter in equilibrium at temperature values relatively low that its contribution can be neglected when computing the equation of state (EoS). In this work we study nuclear matter in the framework of the extended $\sigma$-$\omega$ model with just one variable parameter: the Landau mass, $m_L$. In order to estimate the effect of the variation of $m_L$, we compute the mass radius relations of neutron stars together with their corresponding tidal deformabilities. The former ones have been recently constrained from the observation of X-ray signals from emissions coming from a star surface carried out by the NICER detector~\cite{Miller:2019cac,Riley:2019yda}. The latter ones are associated with the gravitational wave signals of compact stars mergers detected by interferometers of the LIGO-Virgo collaboration~\cite{TheLIGOScientific:2017qsa}.
One crucial question on the neutron star studies is the value of the maximum mass which is quantity dependent on the equation of state (EoS). The stiffer the EoS the larger the maximum mass as well as the stellar radius.  The measurement of the most massive compact star has been carried out by means of the Shapiro delay effect in a binary star system~\cite{Cromartie:2019kug}.
We proceed in our study by considering the constrains from the above observations including their uncertainty ranges to perform a Bayesian analysis in order to obtain posterior probabilities for the Landau mass values.
\iffalse
This paper is organised as follows. In Section~\ref{sec:eos} we describe the equation of state including the parameter values fulfilling empirical data of nuclear quantities. In Section~\ref{sec:bay} we present the neutron star properties derived from the EoS model and show the results from the Bayesian Analysis. We conclude with an outlook for future work in the last part in Section~\ref{sec:sum}.
\fi

%%%%%%%%%%%%%%%%%%%%%%%%%%%%%%%%%%%%%%%%%%%%%%%%%%%%%%%%%%%%%%%%%%%%%%%%%%%%%%%%%%%%%%%%%%%%%%%%%%
\section{The Equation of State}
\label{sec:eos}

For the interior of a compact star we consider here the extended $\sigma $-$\omega$ model which describes protons, electron, and neutrons in $\beta$-equilibrium and approximates the nuclear force by introducing the $\sigma$, $\omega$ and $\rho$ meson. The Lagrange-function corresponding to the extended $\sigma$-$\omega$  model has the following form, 
%
%==================equation===================
\begin{eqnarray}
%\begin{split}
\Lagr &=&
%
%nucleon term
 \overline{\Psi} \left(
i \slashed{\partial} -m_{N} + g_{\sigma} \sigma -g_{\omega} \slashed{\omega}  + g_{\rho} \slashed{\rho}^{a} \tau_{a}
 \right) \Psi
% electron terms
+ \overline{\Psi}_{e} \left(
i \slashed{\partial} - m_{e}
\right) \Psi_{e}
%
%boson term
 +\frac{1}{2}\,\sigma \left(\partial^{2}-m_{\sigma}^2 \right) \sigma - U_{i}(\sigma)  \nonumber \\
%
%vector meson term
&&- \frac{1}{4}\,\omega_{\mu \nu} \omega^{\mu\nu}+\frac{1}{2}m_{\omega}^2 \, \omega^{\mu}\omega_{\mu} 
%
%tensor meson terms
-\frac{1}{4} \rho_{\mu \nu}^{a} \, \rho^{\mu \nu \, a} + \frac{1}{2} m_{\rho}^2 \, \rho_{\mu}^{a} \, \rho^{\mu \, a} 
 \, ,
%\end{split}
\label{eq:wal_lag}
\end{eqnarray}
%===============end of equation================
%
where $\Psi=(\Psi_{n},\Psi_{p})$ is the vector of proton and neutron fields, $m_{N}$ $m_{\sigma}$ $m_{\omega}$ are the $\sigma$ and $\omega$ meson masses and $g_{\sigma}$, $g_{\omega}$, and $g_{\rho}$ are the Yukawa couplings corresponding to the $\sigma$-nucleon,  $\omega$-nucleon and $\rho$-nucleon interactions, respectively. The kinetic terms corresponding to the $\omega$ and $\rho$ meson are written as
%
%==================equation===================
\begin{equation}
%\begin{split}
\omega_{\mu \nu}=\partial_{\mu} \omega_{\nu}-\partial_{\nu} \omega_{\mu} \, , \ \ \textrm{and } \ \
\rho_{\mu \nu}^{a}=\partial_{\mu} \rho_{\nu}^{a} - \partial_{\nu} \rho_{\mu}^{a} + g_{rho} \epsilon^{abc} \rho_{\mu}^{b} \rho_{\nu}^{c}. \, 
%\end{split}
\label{eq:wal_lag}
\end{equation}
%===============end of equation================
%
In eq.~\eqref{eq:wal_lag} $U_{i}(\sigma)$ is a self-interaction term for the $\sigma$-meson and it has the following form:
%
%==================equation list===================
\begin{equation}
\begin{aligned}
U_{34}(\sigma) &=\lambda_{3} \sigma^{3} + \lambda_{4} \sigma^{4}  \, .
\end{aligned}
\label{eq:U_types}
\end{equation}
%===============end of equation list================
%
The model is considered in the mean field approximation at zero temperature and finite chemical potential. These assumptions simplify eq.~\eqref{eq:wal_lag} and all of the kinetic terms become zero and only the following components of the mesons has non-zero value: $\omega_{0}=\omega$ and $\rho_{0}^{3}=\rho$. Using these assumptions the free energy corresponding to the model can be calculated as it is described for example in Ref.~\cite{jakovac2015resummation}: 
%
%==================equation===================
\begin{eqnarray}
%\begin{split}
f_{T} &=&
%
%nucleon term
 f_{F}
\left(
m_{N}-g_{\sigma} \sigma,
\mu_{p} - g_{\omega} \omega + g_{\rho} \rho
\right)
+  f_{F} \left(
m_{N}-g_{\sigma} \sigma,
\mu_{n} - g_{\omega} \omega - g_{\rho} \rho
\right)
+ f_{F} \left(m_{e}, \mu_{e} \right)  \nonumber \\ 
%
%boson term
&&+ \frac{1}{2} m_{\sigma}^{2} \sigma^2  + U_{i}(\sigma)
%
%vector meson term
 - \frac{1}{2} m_{\omega}^2 \omega^2 
%
%tensor meson terms
 - \frac{1}{2} m_{\rho}^2 \rho^2 \, , 
 %\end{split}
\label{eq:wal_f}
\end{eqnarray}
%===============end of equation================
%
where $\mu_{p}$, $\mu_{n}$ and $\mu_{e}$ are the proton, neutron, and electron chemical potential respectively. The $f_{F}$ term describes the free energy contribution corresponding to one fermionic degree of freedom, 
%
%==================equation=================
\begin{equation}
%\begin{split}
f_{F}(T,m,\mu)  = -2 T \int \frac{\dd^3 k}{(2 \pi)^3} 
\ln{\left( 1 + \mathrm{e}^{-\beta \left( E_{k}-\mu \right) } \right)}  \ \ \, 
%\end{split}
\end{equation}
%===============end of equation================
%
where $E_{k}^2  = k^2 + m^2$. In these calculations we used the $T \to 0$ approximation to describe the cold and dense nuclear matter of the compact star, which means that the fermionic free energy has only two variables $f_{F}(m, \mu)$.
The free parameters of the model are determined by using nuclear saturation data~\cite{norman1997compact,meng2016relativistic}. The values used to fit the model are the binding energy $B=-16.3$ MeV, the saturation density, $n_{0} = 0.156$ fm\textsuperscript{-3}, the nucleon effective mass, $m^{*}= 0.6m_{N}$, the nucleon Landau mass $m_{L} =0.83 m_{N}$, compressibility, $K =240$ MeV, and asymmetry energy, $a_{sym} =32.5$ MeV. 
The Landau-mass is given by following Ref.~\cite{norman1997compact}: 
%
%==================equation list===================
\begin{equation}
\begin{aligned}
m_{L} &=\frac{k_{F}}{v_{F}}  \quad \text{with} \quad
v_{F} &=\left.\frac{\partial E_{k}}{\partial k} \right|_{k=k_{F}} \, .
\end{aligned}
\label{eq:landau_mass}
\end{equation}
%===============end of equation list================
%
Where $k=k_{F}$ the Fermi-surface and $E_{k}$ is the dispersion relation of the nucleons. The Landau mass is closely related to the effective nucleon mass in mean field theories: 
%
%==================equation===================
\begin{equation}
\begin{split}
m_{L}= \sqrt{k_{F}^2 + m_{N, eff}^2} \, .
\end{split}
\label{eq:effmass_vs_landau_mass}
\end{equation}
%===============end of equation================
%
Because of this connection the Landau mass and the nucleon effective mass can not be fitted simultaneously \cite{meng2016relativistic}.
In this paper the model is fitted to reproduce the data given above, except for the Landau mass and the effective nucleon mass. These parameters are kept free and determined by comparing the mass radius diagrams corresponding to different values of the Landau mass to neutron star observations.  The compression modulus of the nuclear matter is defined as in Refs.~\cite{Schmitt:2010,norman1997compact}:
%
%==================equation===================
\begin{equation}
\begin{split}
K =k_{F}^2 \frac{\partial^2 }{\partial k_{F}^2} \left( \frac{\epsilon}{n} \right)
= 9 n^2 \frac{\partial^2}{\partial n^2} \left( \frac{\epsilon}{n} \right) \, .
\end{split}
\label{eq:K}
\end{equation}
%===============end of equation================
%
The asymmetry energy has the follwing definition \cite{norman1997compact}: 
%
%==================equation===================
\begin{equation}
a_{sym} = \frac{1}{2} \left. 
\frac{\partial^2 }{\partial t^2} \left( \frac{\epsilon}{n} \right) \right|_{t=0}
\label{eq:asym_def}
\end{equation}
%===============end of equation================
%
where $t=\frac{n_{n}-n_{p}}{n_{B}}$. Its value is fitted as it is described for example in Ref.~\cite{norman1997compact}. For the full neutron star EoS description we complement the extended $\sigma$-$\omega$ model with a low density EoS describing the neutron star crust which is joined at the density value where both EoS models  have equal pressures. The crust EoS used here is SLy4~\cite{Douchin:2000kad}.

%%%%%%%%%%%%%%%%%%%%%%%%%%%%%%%%%%%%%%%%%%%%%%%%%%%%%%%%%%%%%%%%%%%%%%%%%%%%%%%%%%%%%%%%%%%%%%%%%%
\section{Bayesian Analysis Results}
\label{sec:bay}

\begin{figure}	
\includegraphics[width=0.9\textwidth]{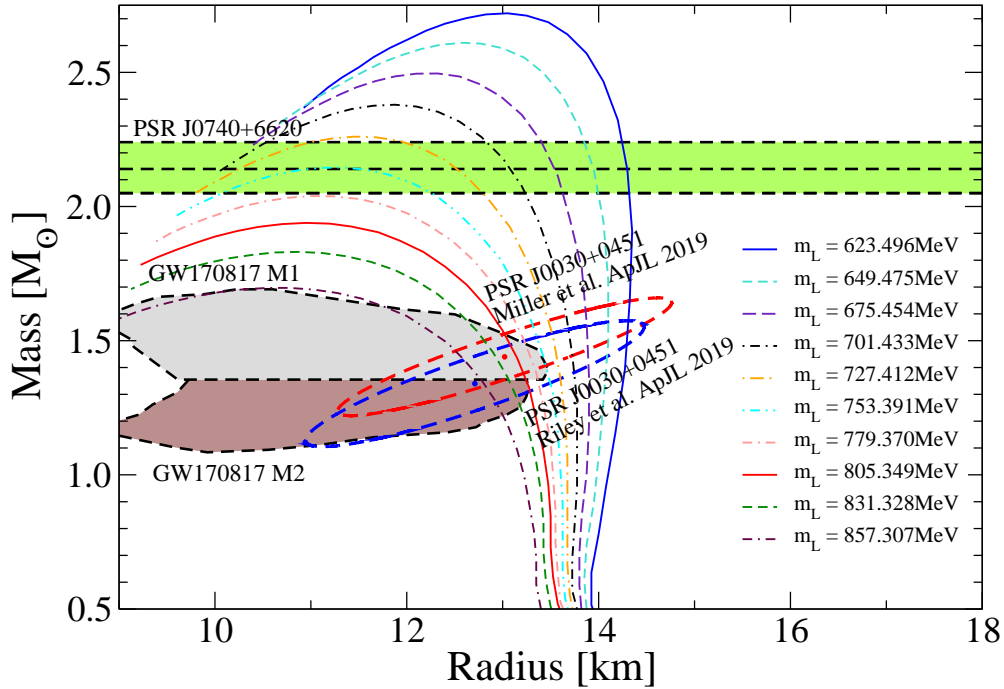}% Here is how to import EPS art
\caption{\label{fig:MvsR} Mass and radius relations of compact stars. Sequences of compact star in the framework of the the extended $\sigma$-$\omega$ model. Each curve corresponds to a value of the Landau mass, $m_L$ which has been treated as a free parameter. The several regions in this diagram denote derived constraints from astrophysical measurements. See the text for details.}
\end{figure}

\begin{figure}[!htb]
\begin{centering}
$\begin{array}{cc}
\includegraphics[width=0.55\textwidth]{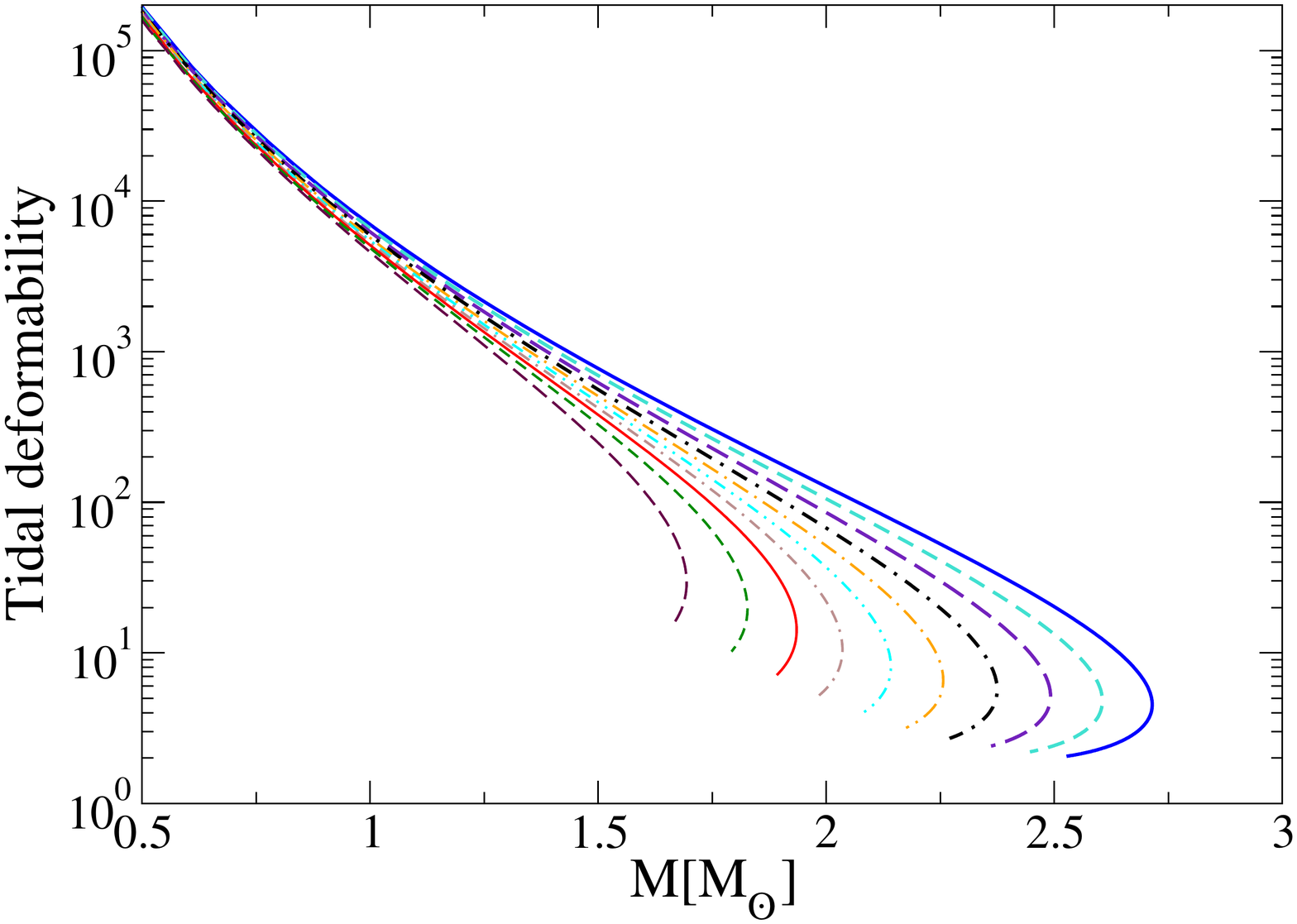} 
& \hspace{-1cm}
\includegraphics[width=0.55\textwidth]{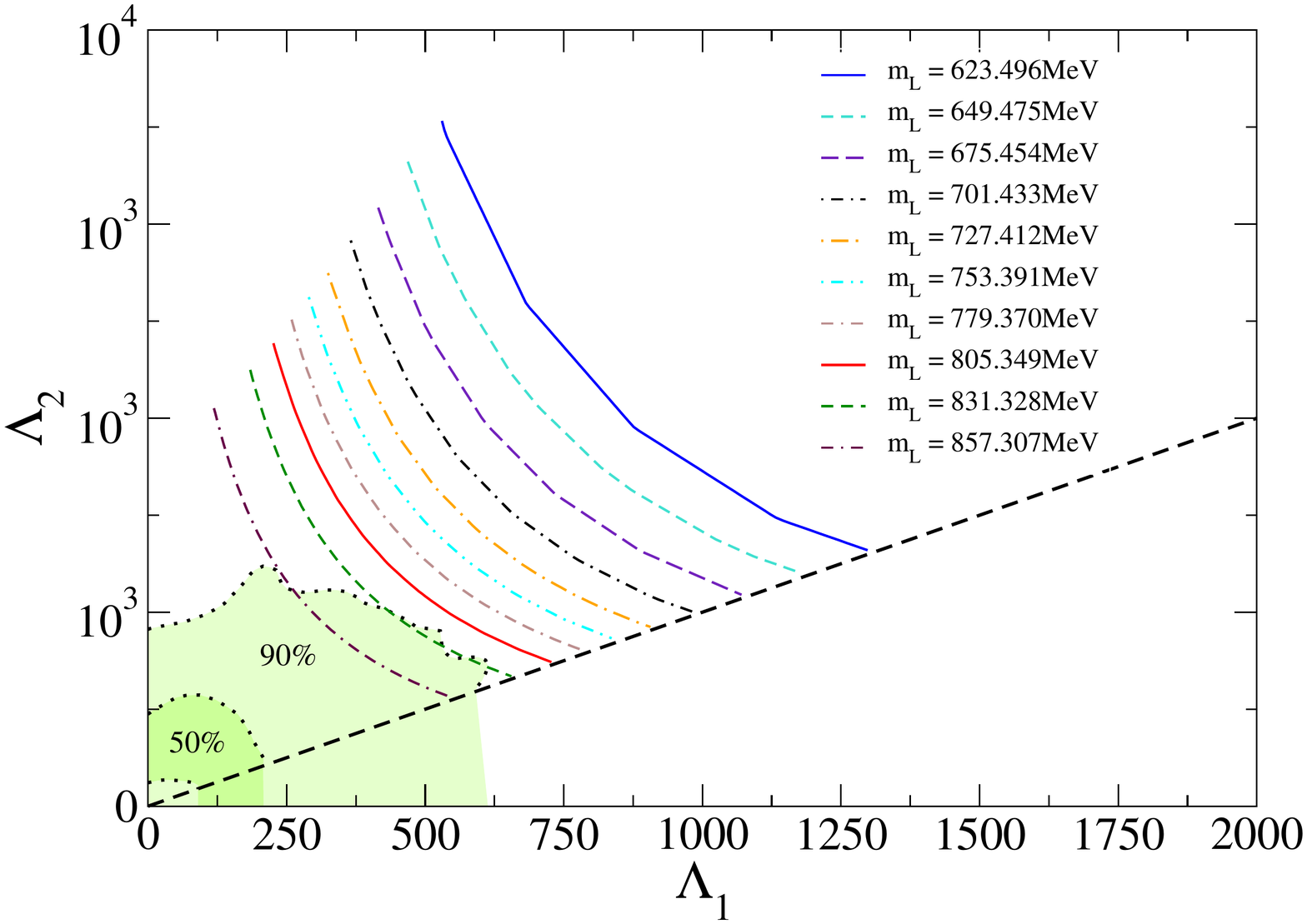}%\vspace{1cm}
\end{array}$ 
\par\end{centering}
\caption{\label{fig:deformability}Tidal deformability $\Lambda$ of compact stars. \textit{Left panel.} $\Lambda$ dependence on the total star mass $M$ for compact star sequences. \textit{Right panel.} Tidal deformabilities derived from the GW170817 event by the LIGO-Virgo collaboration~\cite{TheLIGOScientific:2017qsa}. The green regions represent  $2\sigma$ and $3\sigma$ estimations whereas the different curves correspond to different values of the Landau mass $m_L$ for the EoS model here.}
\end{figure}

The most basic properties of neutron stars are mass and radius. For a static, spherically symmetric neutron star, the solution to the Tolman\,--\,Volkov\,--\,Oppenheimer equations will result into mass-radius relations~\cite{Tolman:1939jz,Oppenheimer:1939ne}. In addition, by considering perturbations to the static metric of the spherical body, it is possible to find its deformability which in the case of inspiraling binaries will produce gravitational waves right before the merger due to tidal forces, see~\cite{Hinderer:2007mb}. Figure \ref{fig:MvsR} shows the resulting compact star sequences together with several regions of constraints from observations. The green, horizontal band centred at $2.14M_{\odot}$ corresponds to the measurement of the most massive pulsar PSR 0740+6620~\cite{Cromartie:2019kug} whereas the gray and brown regions correspond to the estimates of the two components with masses $M_1$ and $M_2$ of the GW170817 merger event. The elliptical dashed regions correspond to the two reported estimates of the PSR J0030+0451 star studied by NICER~\cite{Miller:2019cac,Riley:2019yda}. The results for tidal deformabilities are shown in Figure~\ref{fig:deformability}, in which each line style corresponds to a $m_L$ value. A systematic behaviour on the mass-radius relation is clear: lower (higher) values of $m_L$ correspond to stiffer (softer) neutron star matter and higher (lower) maximum mass. On the $\Lambda_1 - \Lambda_2$ diagram the darker green region signifies a $2\sigma$ confidence level whereas the lighter ones correspond to $3\sigma$.
Next, we perform a Bayesian inference study. Many other works~\cite{Alvarez-Castillo:2016oln,Raithel:2017ity,Ayriyan:2018blj,Miller:2019nzo,Raaijmakers:2019dks}  have applied this kind of analysis basing their result on other constraints usually bearing higher uncertainties than the ones we have chosen,  for example a primer work on X-ray bursts analysis from the surface of compact stars~\cite{Steiner:2010fz}. Instead, in our work we consider the recent NICER measurements for a mass-radius estimate of the pulsar PSR J0030+0451 together with the measurement of the mass of the most massive observed pulsar PSR 0740+6620 and the information derived from the gravitational wave detection event GW170817. 
%

%%%%%%%%%%%%%%%%%%%%%%%%%%%%%%%%%%%%%%%%%%%%%%%%%%%%%%%%%%%%%%%%%%%%%%%%%%%%%%%%%%%%%%%%%%%%%%%%%%
\section{Outlook and Conclusions}
\label{sec:sum}

\begin{figure*}[!htb]
\begin{centering}
$\begin{array}{cc}
\label{Deformability}
\includegraphics[width=0.5\textwidth]{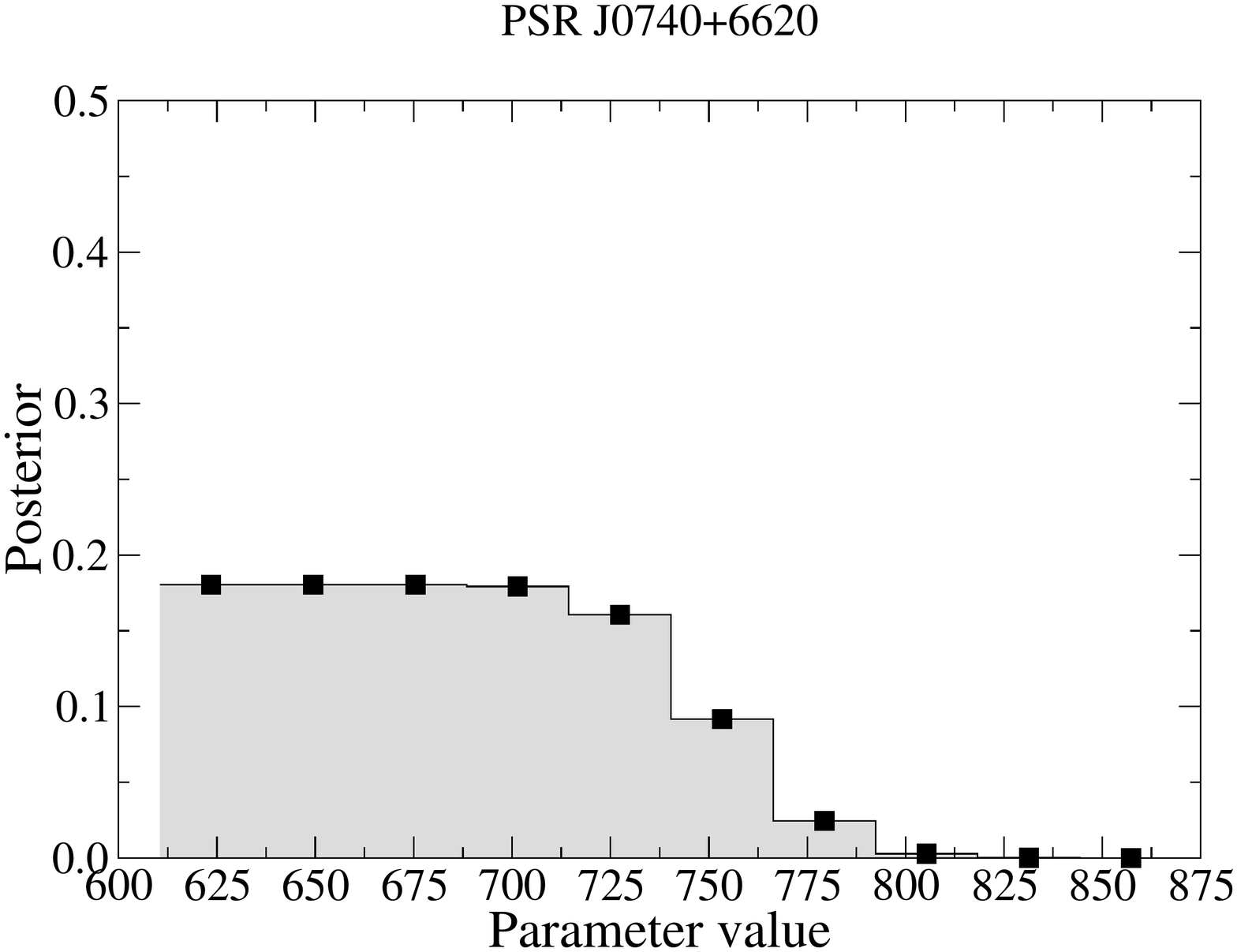}  & \hspace{0cm} \includegraphics[width=0.5\textwidth]{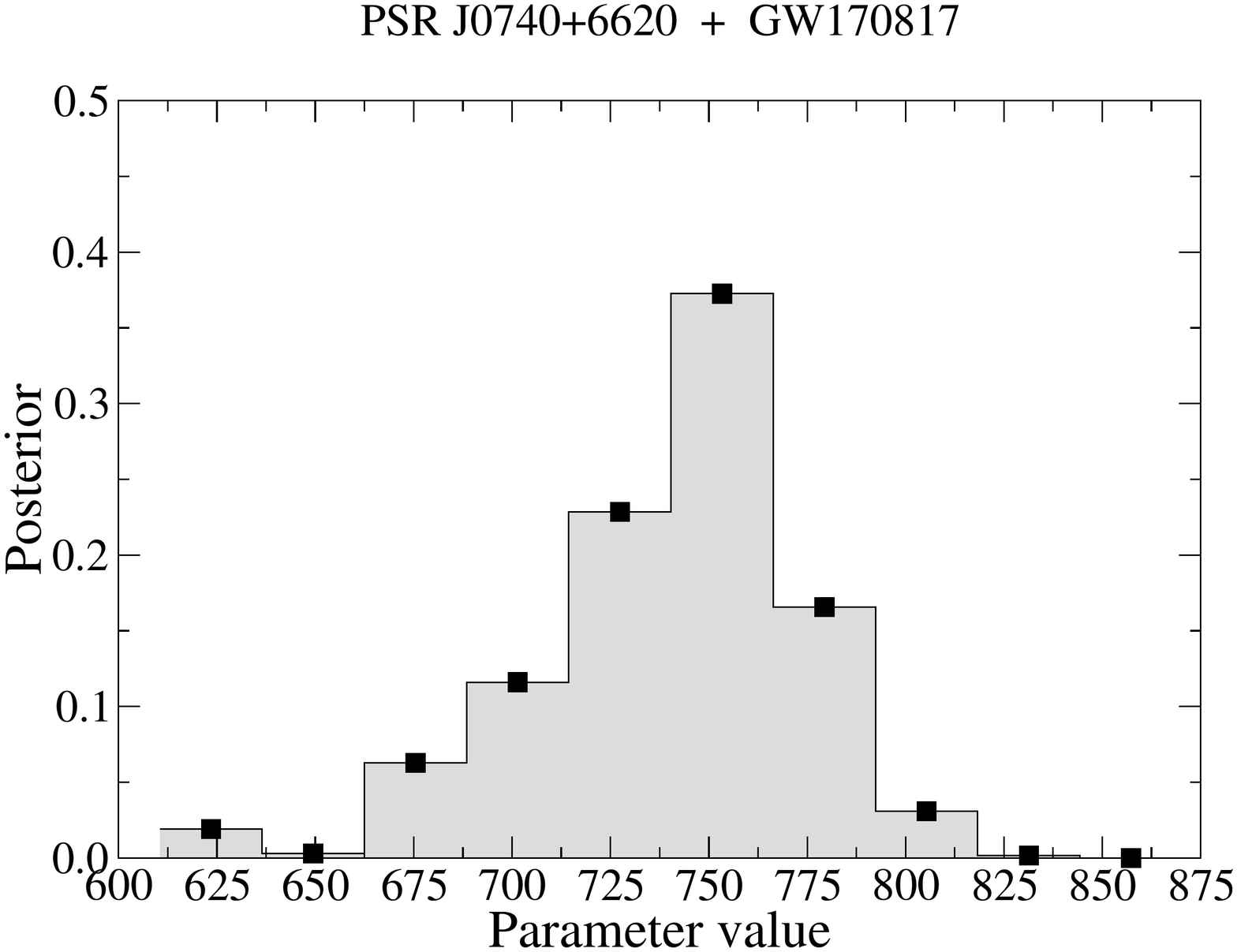}\\
\includegraphics[width=0.5\textwidth]{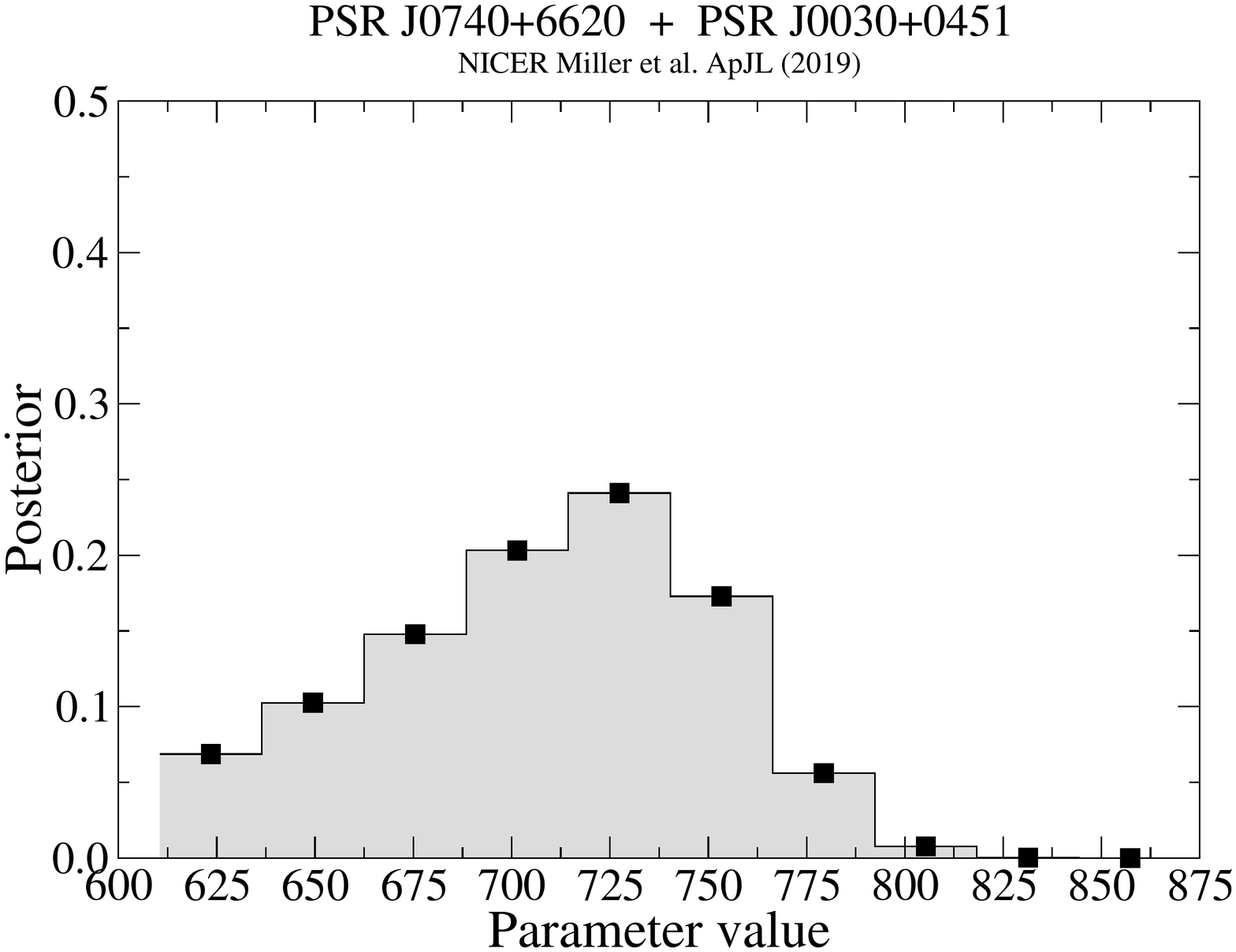}  & \hspace{0cm} \includegraphics[width=0.5\textwidth]{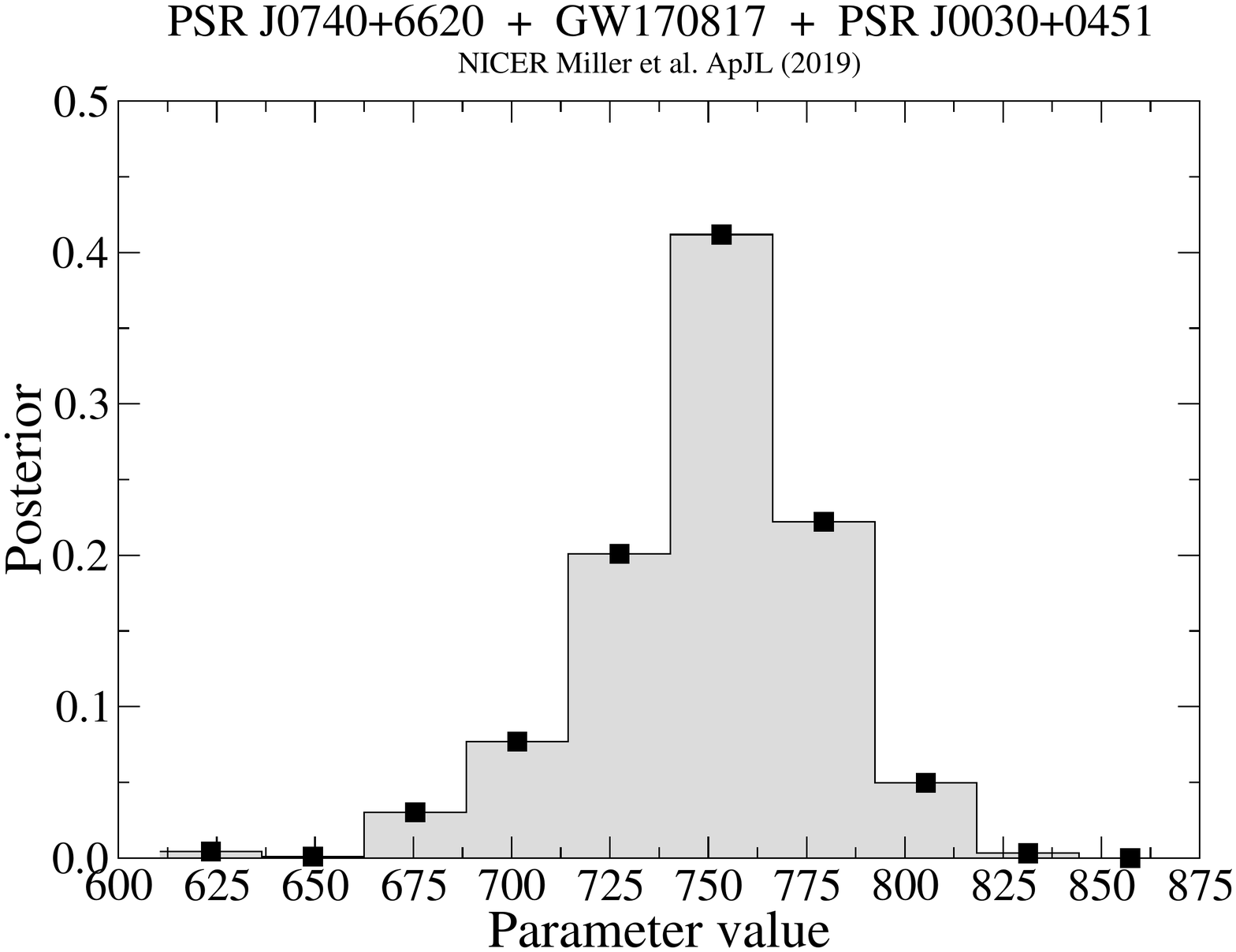}\\
\includegraphics[width=0.5\textwidth]{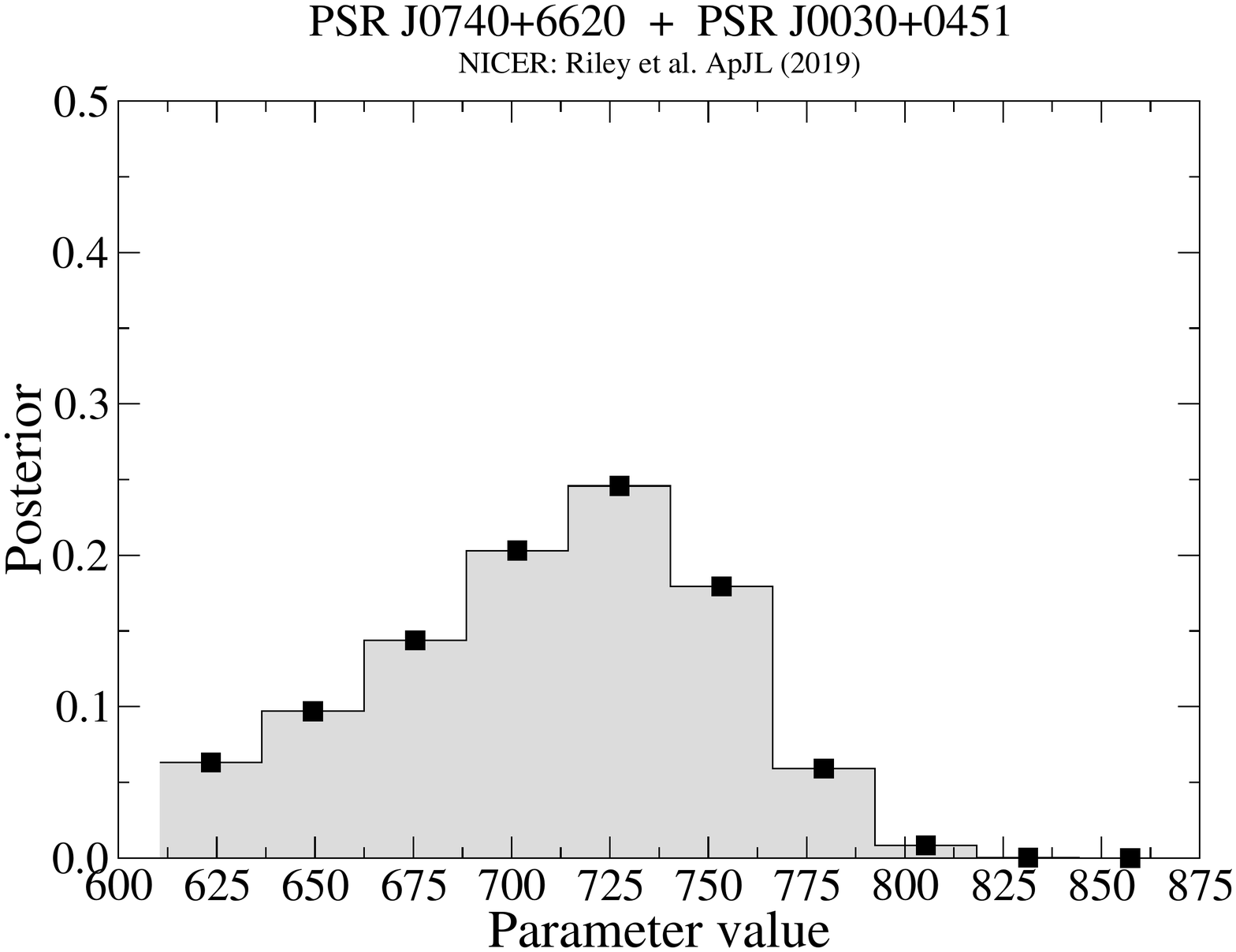}  & \hspace{0cm} \includegraphics[width=0.5\textwidth]{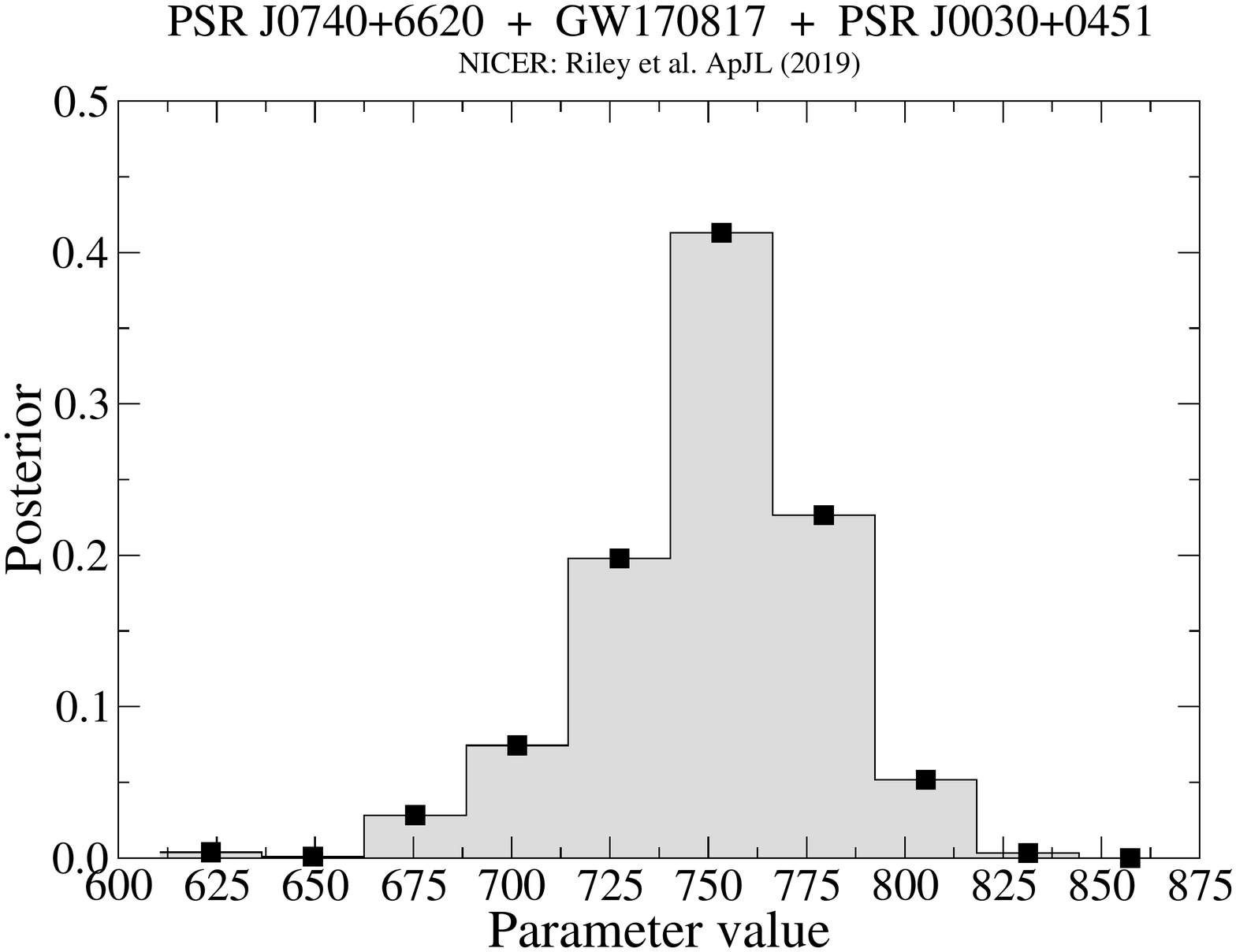}\\
\end{array}$ 
\par\end{centering}
\caption{\label{posteriors} Posterior probabilities of the Landau mass $m_L$ parameter resulting from a Bayesian analysis. This set of figures show how the probabilities change depending on the measurements considered. In the case of the PSR J0030+0451 mass and radius values by NICER, we consider both reported values by Miller et al.~\cite{Miller:2019cac} and Riley et al.~\cite{Riley:2019yda} and we found no significant difference in the results when choosing either of them.}
\end{figure*}

The posterior probability distributions of our Bayesian analysis are shown in Figure~\ref{posteriors}, which displays 6 panels in which compact star observations have been considered. For the full analysis that includes three measurements, the probabilities peak at a Landau mass value around $m_L=750\pm 15$~MeV. As it can be seen from the figure, the choice of either one of the NICER measurement values does not significantly affect the resulting probabilities. Incorporating new constraints, like the recently reported second compact star merger GW190425~\cite{Abbott:2020uma}
shall be subject of a follow-up work. Moreover, a variation of multiple parameters of the model  may allow for an improvement of the probabilities to better fulfill both laboratory empirical values as well as compact star observations.

%%%%%%%%%%%%%%%%%%%%%%%%%%%%%%%%%%%%%%%%%%%%%%%%%%%%%%%%%%%%%%%%%%%%%%%%%%%%%%%%%%%%%%%%%%%%%%%%%%
\begin{acknowledgments}
Authors GGB and PP was supported by Hungarian National Research Fund (OTKA) grants K120660, K123815, NKFIH 2019-2.1.11-TÉT-2019-00050, 2019-2.1.11-TÉT-2019-00078, COST actions CA15213 (THOR) and CA16214 (PHAROS). A. A. was supported by the RFBR grant No. 18-02-40137. Authors also acknowledge the computational resources for the Wigner GPU Laboratory.
\end{acknowledgments}

%%%%%%%%%%%%%%%%%%%%%%%%%%%%%%%%%%%%%%%%%%%%%%%%%%%%%%%%%%%%%%%%%%%%%%%%%%%%%%%%%%%%%%%%%%%%%%%%%%

\end{document}